\documentclass[aps,pra,a4paper,twocolumn,showpacs]{revtex4}
\usepackage[dvips]{epsfig}
\usepackage{subfigure}
\usepackage{amssymb}
\usepackage{amsmath}
\usepackage[usenames]{color}
\usepackage{pstricks}   

\begin{document}

\preprint{JChemPhys}

\title{Quantum calculations of H$_2$--H$_2$ collisions: from ultracold to thermal energies}

\author{Goulven Qu{\'e}m{\'e}ner, Naduvalath Balakrishnan}
\affiliation{Department of Chemistry, University of Nevada Las Vegas,
Las Vegas, NV 89154, United States of America}

\date{\today}

\begin{abstract}
We present  quantum dynamics
of collisions between two para-H$_2$ molecules
from low (10$^{-3}$~K)
to high collision energies (1~eV). The calculations are carried out using a 
quantum scattering code that solves the time-independent Schr\"odinger equation in its
full dimensionality without any decoupling approximations. 
The six-dimensional potential energy surface for the H$_4$ system developed by
Boothroyd et al. [J. Chem. Phys. {\bf 116}, 666 (2002)] is used in the calculations.
Elastic, inelastic and state-to-state cross sections
as well as rate coefficients from $T = 1$~K to 400~K obtained from our calculations are compared 
with available experimental and theoretical results. Overall,
good agreement is obtained with previous studies.
\end{abstract}

\maketitle

\font\smallfont=cmr7

\section{Introduction}

Collisions between molecules are fundamental processes which
take place 
in many areas of physics and chemistry.
Molecular collisions are prevalent in the interstellar medium,
in the atmospheres of planets, in combustion chemistry and 
and in many other chemical and industrial processes.
Being the simplest neutral molecule - molecule system, the H$_2-$H$_2$ system
has served as a prototype 
for accurate calculations of tetratomic potential energy surfaces (PESs)
as well as accurate quantum dynamics treatment of diatom - diatom collisions.
Since H$_2$ is the most abundant molecular species
in the universe, 
collisions between H$_2$ molecules leading to excitation and de-excitation of their
rovibrational levels continue to be a topic of considerable interest in interstellar
chemistry. However, an accurate calculation of rovibrational energy
transfer rate coefficients in H$_2-$H$_2$ collisions for temperatures relevant
to astrophysical environments is still a challenging problem. In recent years,
due to the success in creating dense samples of ultracold molecules in magneto-optical
traps, atomic and molecular collisions in cold and ultracold gases have attracted
considerable experimental and theoretical 
attention~\cite{Bahns00,Masnou01,Bethlem03,Doyle04,Hutson06,Hutson07a,roman-review,Roman-pccp}. 
These unusual systems provide a 
fascinating opportunity to investigate atomic and molecular processes at temperatures
close to absolute zero where the collisional outcomes are dramatically influenced
by quantum effects. Recent studies of atom - diatom inelastic and reactive collisions at 
temperatures close to absolute zero have shown that such processes may occur with
significant rate coefficients at ultracold 
temperatures~\cite{bala-cpl-2001,weck06,Quemener08b,bodo-review,Soldan02,Quemener04,Quemener05,Cvitas05a,Cvitas05b,Quemener07,Quemener09}.
The H$_2-$H$_2$ system serves 
as an ideal prototype to investigate molecule - molecule collisions at ultracold
temperatures.

Several PESs of the H$_4$ system
have been calculated in the literature.
Some of the earlier studies treated the diatomic molecules as a rigid rotor (RR)
as in the work of 
Zarur and Rabitz (ZR)~\cite{Zarur74},
Schaefer and K\"ohler (SK)~\cite{Schaefer89},
Diep and Johnson (DJ)~\cite{Diep00},
and the recent study of Patkowski et al.~\cite{Patkowski08}. Some of these PESs
have been employed in several studies of rotational energy transfer in H$_2-$H$_2$
collisions within the rigid-rotor approximation at thermal 
energies~\cite{Zarur74,Forrey01,Forrey02,Mate05,Lee06,Sultanov06a,Sultanov06b}
and also at ultralow energies~\cite{Forrey01,Forrey02,Lee06}.
Several full-dimensional potential surfaces have also been reported for the H$_4$ system
in recent years. They include
the PESs of Schwenke~\cite{Schwenke88},
Aguado, Su{\'a}rez and Paniagua (ASP)~\cite{Aguado94},
Boothroyd, Martin, Keogh and Peterson (BMKP)~\cite{Boothroyd02},
and the more recent work of Hinde~\cite{Hinde08}.
These PESs have been adopted in a number of  quantum 
scattering calculations of H$_2-$H$_2$ that go beyond the rigid rotor 
approximation~\cite{Flower98a,Flower98b,Flower99,Flower00a,Flower00b}.
The  BMKP and Schwenke PESs were recently employed in full-dimensional quantum calculations by 
Pogrebyna et al.~\cite{Pogrebnya02,Pogrebnya03}.
They reported the first quantum calculation of vibrational relaxation
in collisions between H$_2(v=1,j=0)$ and H$_2(v=0,j=0)$ molecules
using a time-independent quantum mechanical (TIQM)
method within an angular momentum decoupling approximation
called the coupled-states approximation (CSA).
Lin et al.~\cite{Lin02} used a time-dependent wave packet (TDWP) method
and the CSA method to study pure rotational transitions in H$_2(v=0,j=0)$ + H$_2(v=0,j=0)$ collisions
at thermal energies.
Full-dimensional quantum 
studies without CSA have also been recently reported using the TDWP methods~\cite{Gatti05,Otto08,Panda07}. 
Gatti et al.~\cite{Gatti05} and Otto et al.~\cite{Otto08}
investigated para-para collisions involving
H$_2(v=0,j=0)$ + H$_2(v=0,j=0)$,
and Panda et al.~\cite{Panda07} 
reported ortho-para collisions 
of H$_2(v=1,0,j=1)$ + H$_2(v=0,1,j=0)$. 
In a recent Communication~\cite{Quemener08a}, 
we  reported rotational and 
vibrational relaxation of H$_2(v=1,j=0,2)$ in collisions with H$_2(v=0,j=0,2)$
at ultralow energies
using the BMKP surface and a TIQM method that does not involve 
any angular momentum decoupling approximation.
The calculations showed that indistinguishable molecule - molecule collisions may involve highly
efficient near-resonant energy transfer if they are accompanied by simultaneous conservation of
the total rotational angular momentum and internal energy of the molecules.
This mechanism was found to be independent of the initial vibrational excitation of the molecules.

In this article, we present the full 
quantum dynamics of rovibrational energy transfer in para-H$_2$ + para-H$_2$ system
from ultralow to high collision energies and compare our 
results with previous quantum calculations~\cite{Pogrebnya02,Lee06,Forrey01,Otto08}
and available
experimental results~\cite{Audibert75,Bauer76,Mate05}.
We employ the  BMKP PES and the 
full-dimensional TwoBC - quantum scattering program developed by Krems~\cite{Roman06}
implemented in our previous study~\cite{Quemener08a}. 
The paper is organized as follows: In section II,
we give an overview of the TIQM scattering theory of two $^1\Sigma$ diatomic molecules.
In section III, 
we present state-to-state and initial state-selected cross sections and rate coefficients
for the 
H$_2(v=1,j=0)$ + H$_2(v=0,j=0)$ 
and H$_2(v=0,j=0)$ + H$_2(v=0,j=0)$ collisions.
Conclusions are given in section IV.

\section{Molecule - molecule scattering theory}

The TIQM formalism for scattering of  two $^1\Sigma$ diatomic molecules has been
described by Takayanagi~\cite{Takayanagi65}, Green~\cite{Green75}, 
Alexander and DePristo~\cite{Alexander77}, and Zarur and Rabitz~\cite{Zarur74},
using
the close-coupling (CC) formalism of Arthurs and Dalgarno~\cite{Arthurs60}.
A brief account of the theoretical formalism with particular emphasis on
distinguishable versus indistinguishable molecule collisions is given here.
For more details we refer to the original papers.

\begin{figure}[h]
\begin{center}
\includegraphics*[bb=180 50 542 892,height=10cm,keepaspectratio=true,angle=-90]{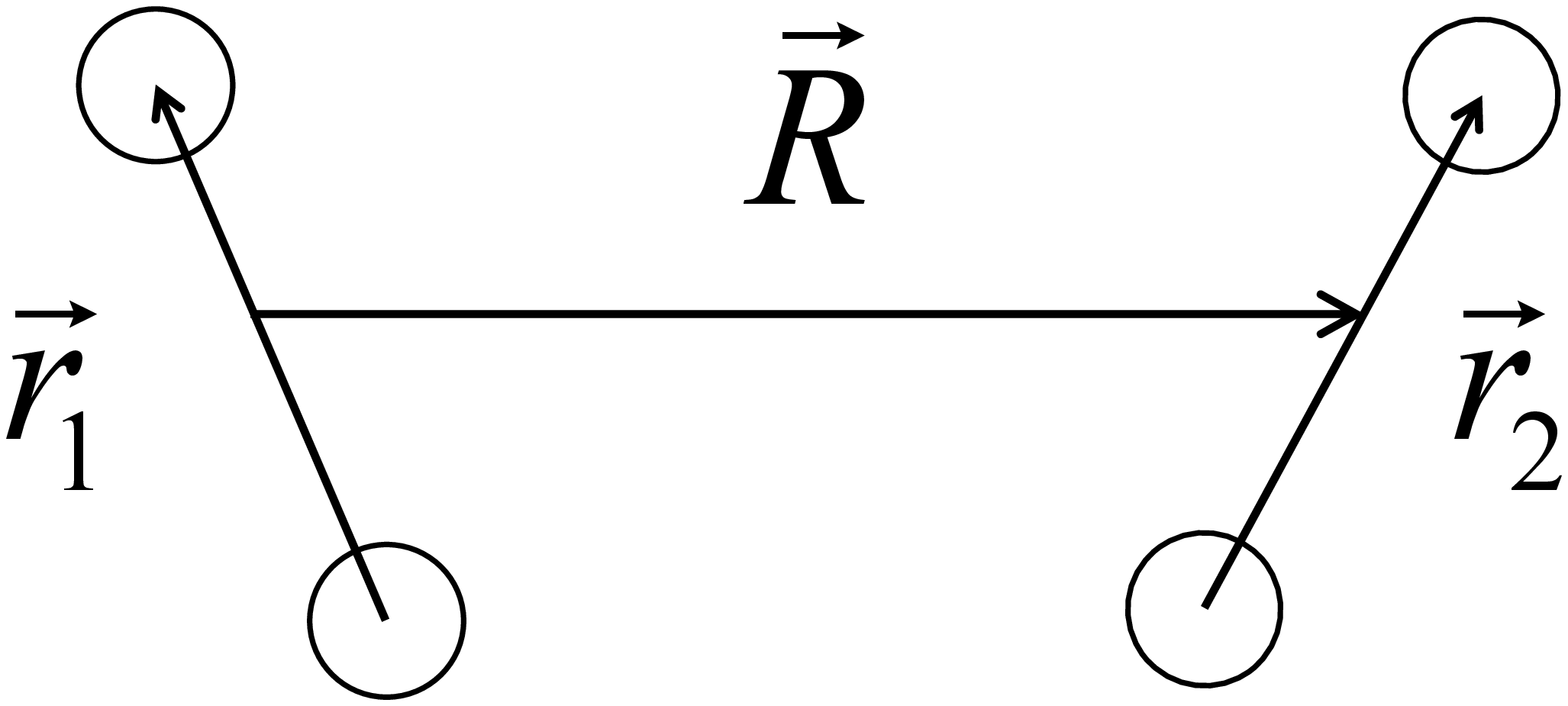}
\caption{
Jacobi vectors used to describe molecule - molecule collisions.
\label{JacCord-FIG}}
\end{center}
\end{figure}

Figure~\ref{JacCord-FIG} shows the Jacobi vectors employed in the calculations where
$\vec{r}_1(r_1,\hat{r}_1)$ and $\vec{r}_2(r_2,\hat{r}_2)$ describe the vectors
joining the two atoms of the two H$_2$ molecules and  $\vec{R}(R,\hat{R})$
denote the 
vector joining the centers of mass of the two molecules.
The Hamiltonian of the system,
\begin{eqnarray}
H(\vec{r}_1,\vec{r}_2,\vec{R}) = T(\vec{r}_1) + T(\vec{r}_2) + T(\vec{R}) + V(\vec{r}_1,\vec{r}_2,\vec{R}) ,
\label{HAM}
\end{eqnarray}
is composed of a radial kinetic energy term
$T(\vec{R})$ describing the center-of-mass motion,  
two kinetic energy terms $T(\vec{r}_1)$ and $T(\vec{r}_2)$
for each diatomic molecule and
the PES function,
\begin{eqnarray}
V(\vec{r}_1,\vec{r}_2,\vec{R}) = U(\vec{r}_1,\vec{r}_2,\vec{R}) + V(\vec{r}_1) + V(\vec{r}_2) ,
\label{Potential}
\end{eqnarray}
describing the interactions between the four atoms.
The terms $V(\vec{r}_1)$ and $V(\vec{r}_2)$ in Eq.~\eqref{Potential} 
are the interaction energy potentials
of the two H$_2$ molecules and
$U(\vec{r}_1,\vec{r}_2,\vec{R})$ is the interaction energy potential 
between the two molecules, which
vanishes at large molecule - molecule separations,
\begin{eqnarray}
\lim_{R\to\infty}^{} U(\vec{r}_1,\vec{r}_2,\vec{R}) \to 0. 
\end{eqnarray}
The angular dependence of the interaction potential may be expanded as
\begin{eqnarray}
U(\vec{r}_1,\vec{r}_2,\vec{R}) = \sum^{}_{\lambda} \ 
A_{\lambda}(r_1,r_2,R)  Y_{\lambda}(\hat{r}_1,\hat{r}_2,\hat{R}) ,
\label{POTLAMBDA}
\end{eqnarray}
with
\begin{multline}
Y_{\lambda}(\hat{r}_1,\hat{r}_2,\hat{R}) = 
\sum^{}_{m_{\lambda}} \big\langle  \lambda_1 m_{\lambda_1} \lambda_2 m_{\lambda_2} \big| \lambda_{12} m_{\lambda_{12}} \big\rangle \\
Y_{\lambda_1 m_{\lambda_1}}(\hat{r}_1) Y_{\lambda_2 m_{\lambda_2}}(\hat{r}_2) Y^*_{\lambda_{12} m_{\lambda_{12}}}(\hat{R}) ,
\end{multline}
where $\lambda \equiv \lambda_1 \lambda_2 \lambda_{12}$ and $m_{\lambda} \equiv m_{\lambda_1} m_{\lambda_2} m_{\lambda_{12}}$.
The notation $\big\langle  \lambda_1 m_{\lambda_1} \lambda_2 m_{\lambda_2} \big| \lambda_{12} m_{\lambda_{12}} \big\rangle$ 
represents a Clebsch--Gordan coefficient.
Because  the Hamiltonian defined in Eq.~\eqref{HAM} is invariant under a  rotation
in space, 
the total angular momentum $\vec J$ and its projection $J_z$ on a space-fixed axis
are conserved during the collision. 
In the
following, we will discuss the close-coupling equations for both
distinguishable and indistinguishable molecule - molecule collisions.

For collisions between two distinguishable molecules,
%
the diabatic functions,
\begin{eqnarray}
\phi^{JM\varepsilon_I}_{vjl}(\vec{r}_1,\vec{r}_2,\hat{R}) 
= \chi_{vj_1j_2}(r_1,r_2) \big\langle \hat{r}_1 \hat{r}_2 \hat{R} \big| j l J M \big\rangle ,
\label{diabfuncbasis}
\end{eqnarray}
serve as basis functions to expand the total
wave function,
%
\begin{multline}
\Psi(\vec{r}_1,\vec{r}_2,\vec{R}) = \\
\frac{1}{R} 
\sum_{v,j,l,J,M}^{} F_{vjl}^{JM\varepsilon_I}(R) \phi^{JM\varepsilon_I}_{vjl}(\vec{r}_1,\vec{r}_2,\hat{R}) , 
\label{PSI}
\end{multline}
%
where $v \equiv v_1 v_2$ denotes the vibrational quantum numbers
and $j \equiv j_1 j_2 j_{12}$ denotes the rotational quantum numbers of the molecules.
The function $\chi_{vj_1j_2}(r_1,r_2) = \chi_{v_1j_1}(r_1) \chi_{v_2j_2}(r_2)$
is the product of the vibrational wave functions of the two diatomic fragments and the 
angular function,
\begin{multline}
\langle \hat{r}_1 \hat{r}_2 \hat{R}  \big| j l J M \rangle =  \\
\sum^{}_{m_{j_1}, m_{j_2}, m_{j_{12}}, m_l} \langle  j_1 m_{j_1} j_2 m_{j_2} \big| j_{12} m_{j_{12}} \rangle  
\big\langle j_{12} m_{j_{12}} l m_l \big| J M \big\rangle \\
Y_{j_1 m_{j_1}}(\hat{r}_1) Y_{j_2 m_{j_2}}(\hat{r}_2) Y_{l m_l}(\hat{R}),
\end{multline}
denotes the rotational 
wave function of the two molecules in the total angular momentum representation. 
In this representation,
the rotational angular momenta
$\vec{j}_1$ and $\vec{j}_2$
are coupled to give the total rotational momentum $\vec{j}_{12}$ which is subsequently 
coupled to the orbital angular momentum $\vec{l}$
to yield the total angular momentum $\vec{J}$.
Under spatial inversion $I$, the basis functions~\eqref{diabfuncbasis} obey the eigenvalue  equation,
\begin{eqnarray}
I \phi^{JM\varepsilon_I}_{vjl} = \varepsilon_I \phi^{JM\varepsilon_I}_{vjl} ,
\end{eqnarray}
with $\varepsilon_I  = (-1)^{j_1+j_2+l} = \pm 1$.
Substitution of Eq.~\eqref{HAM} and Eq.~\eqref{PSI}
in the time-independent Schr{\"o}dinger equation $H \Psi = E \Psi$
leads to a set of close-coupling equations in the radial coordinate, $R$. 
%
%
%
The coupled equations become
\begin{multline}
\left\{ - \frac{\hbar^2}{2 \mu} \frac{d^2}{d R^2} +
\frac{\hbar^2 l(l+1)}{2 \mu R^2} 
+ \varepsilon_{vj} - E \right\}
\ F_{vjl}^{JM\varepsilon_I}(R)  \\
+ \sum_{v'j'l'}
  {\cal U}^{JM\varepsilon_I}_{vjl,v'j'l'}(R)
\ F_{v'j'l'}^{JM\varepsilon_I}(R) = 0,
\label{CC1}
\end{multline}
where $\mu$ is the reduced mass of the molecule - molecule system.
The total energy
is $E=\varepsilon_{vj}+E_c$
where $\varepsilon_{vj} = \varepsilon_{v_1j_1} + \varepsilon_{v_2j_2}$
is the rovibrational energy of the two separated molecular fragments
and $E_c$ is the collision energy.
The interaction potential matrix is given by
\begin{eqnarray}
{\cal U}^{JM\varepsilon_I}_{vjl,v'j'l'}(R) = 
\sum_{\lambda} B^{\lambda}_{vj_1j_2,v'j'_1j'_2}(R) f^{J;\lambda}_{jl,j'l'} . 
\end{eqnarray}
The radial interaction potential term $B^{\lambda}_{vj_1j_2,v'j'_1j'_2}(R)$ 
is given by 
\begin{multline}
B^{\lambda}_{vj_1j_2,v'j'_1j'_2}(R) =
\int_0^{\infty} \int_0^{\infty} \chi_{vj_1j_2}(r_1,r_2)  \\
A_{\lambda}(r_1,r_2,R)  \chi_{v'j'_1j'_2}(r_1,r_2) dr_1 dr_2  ,
\end{multline}
and the  function $f^{J;\lambda}_{jl,j'l'}$
is given in terms of $3-j$, $6-j$, and $9-j$ symbols by
\begin{multline}
f^{J;\lambda}_{jl,j'l'} = 
(4\pi)^{-3/2} (-1)^{j_1+j_2+j'_{12}+J}[\lambda,j,l,j',l',\lambda_{12}]^{1/2} \\
\left( \begin{array}{ccc} j_1 & j'_1 & \lambda_1 \\ 0 & 0 & 0 \end{array} \right)
\left( \begin{array}{ccc} j_2 & j'_2 & \lambda_2 \\ 0 & 0 & 0 \end{array} \right)
\left( \begin{array}{ccc} l & l' & \lambda_{12} \\ 0 & 0 & 0 \end{array} \right) \\
\left\{ \begin{array}{ccc} l & l' & \lambda_{12} \\ j'_{12} & j_{12} & J \end{array} \right\}
\left\{ \begin{array}{ccc} j'_{12} & j'_2 & j'_1 
\\ j_{12} & j_2 & j_1 \\ \lambda_{12} & \lambda_2 & \lambda_1  \end{array} \right\}, 
\end{multline}
with the notation
\begin{eqnarray}
[x_1, x_2, ... , x_n] = (2x_1+1) (2x_2+1) ... (2x_n+1).
\end{eqnarray}

For two indistinguishable molecules, one has to symmetrize
the wave function with respect to the exchange-permutation symmetry of the molecules.
Under molecule permutation $P$,
$P \phi^{JM\varepsilon_I}_{vjl} = (-1)^{j_1+j_2+j_{12}+l} \phi^{JM\varepsilon_I}_{\bar{v}\bar{j}l}$,
where $\bar{v} \equiv v_2v_1$ and $\bar{j} \equiv j_2j_1j_{12}$. 
As a consequence,  properly symmetrized exchange-permutation invariant states must be
employed for the basis set,
\begin{multline}
\phi^{JM\varepsilon_I\varepsilon_P}_{vjl}(\vec{r}_1,\vec{r}_2,\hat{R}) = \\
\Delta_{vj_1j_2} 
\left[ \phi^{JM\varepsilon_I}_{vjl}
+ \varepsilon_P (-1)^{j_1+j_2+j_{12}+l} 
\phi^{JM\varepsilon_I}_{\bar{v}\bar{j}l} \right] ,
\end{multline}
with $\varepsilon_P = \pm 1$, $\Delta_{vj_1j_2}=[2(1+\delta_{v_1v_2}\delta_{j_1j_2})]^{-1/2}$,
and
\begin{eqnarray} 
P \phi^{JM\varepsilon_I\varepsilon_P}_{vjl} = \varepsilon_P \phi^{JM\varepsilon_I\varepsilon_P}_{vjl}. 
\end{eqnarray} 
The total wavefunction is expanded as
\begin{multline}
\Psi(\vec{r}_1,\vec{r}_2,\vec{R}) = \\
\frac{1}{R}
\sum_{v,j,l,J,M}^{*} F_{vjl}^{JM\varepsilon_I\varepsilon_P}(R) 
\phi^{JM\varepsilon_I\varepsilon_P}_{vjl}(\vec{r}_1,\vec{r}_2,\hat{R}),
\end{multline}
%
where the superscript over the sum symbol 
indicates a summation over states
that have to follow a ``well ordered'' classification
as discussed in Ref.~\cite{Takayanagi65,Green75,Alexander77}.
We choose
$v_1 > v_2$
or when $v_1=v_2$, $j_1 \ge j_2$.
The time-independent Schr{\"o}dinger equation 
yields a set of close-coupling equations, 
\begin{multline}
\left\{ - \frac{\hbar^2}{2 \mu} \frac{d^2}{d R^2} +
\frac{\hbar^2 l(l+1)}{2 \mu R^2}
+ \varepsilon_{vj} - E \right\}
\ F_{vjl}^{JM\varepsilon_I \varepsilon_P}(R)  \\
+ \sum_{v'j'l'}^{*} 
  {\cal U}^{JM\varepsilon_I \varepsilon_P}_{vjl,v'j'l'}(R)
\ F_{v'j'l'}^{JM\varepsilon_I \varepsilon_P}(R) = 0,
\label{CC2}
\end{multline}
with the interaction potential matrix,
\begin{multline}
{\cal U}^{JM\varepsilon_I \varepsilon_P}_{vjl,v'j'l'}(R) = 
2 \Delta_{vj_1j_2} \Delta_{v'j'_1j'_2} [ {\cal U}^{JM\varepsilon_I}_{vjl,v'j'l'}(R) \\
+ \varepsilon_P (-1)^{j'_1+j'_2+j'_{12}+l'} 
{\cal U}^{JM\varepsilon_I}_{vjl,\bar{v}'\bar{j}'l'}(R)].
\end{multline}
Because of the indistinguishability
of the molecules and the summation over the ``well ordered'' states,
the size of the set of close-coupling equations~\eqref{CC2} is 
smaller than the set~\eqref{CC1}.

The close-coupling equations~\eqref{CC1} or~\eqref{CC2}
are solved for each value of $R$
using the log-derivative matrix propagation method of Johnson~\cite{Johnson73}
and Manolopoulos~\cite{Manolopoulos86}.
The log-derivative matrix is propagated
to a matching distance
where asymptotic boundary conditions
are applied to obtain the scattering matrix, $S$,
for given values of $J$, $\varepsilon_I$ and $\varepsilon_P$.
For distinguishable  molecule collisions,
the state-to-state cross section is given by
\begin{multline}
\sigma_{v_1j_1v_2j_2 \to v'_1j'_1v'_2j'_2}(E_c) = \frac{\pi}{(2j_1+1)(2j_2+1)\text{k}^2} \\
\sum_{j_{12}j'_{12}ll'J\varepsilon_I}^{} (2J+1) 
|\delta_{vjl,v'j'l'} 
 - S^{J\varepsilon_I}_{vjl,v'j'l'}(E_c)|^{2},
\end{multline}
where the wave vector $\text{k}^2=2 \mu E_c/\hbar^2$.
For indistinguishable molecules,
the state-to-state cross section  is given by
a statistically weighted sum of the exchange-permutation symmetrized cross sections,
\begin{eqnarray}
\sigma_{v_1j_1v_2j_2 \to v'_1j'_1v'_2j'_2}(E_c)
= W^+ \sigma^{\varepsilon_P=+1} + W^- \sigma^{\varepsilon_P=-1} ,
\end{eqnarray}
where
\begin{multline}
\sigma^{\varepsilon_P}
= \frac{\pi(1+\delta_{v_1v_2}\delta_{j_1j_2})(1+\delta_{v'_1v'_2}\delta_{j'_1j'_2})}
{(2j_1+1)(2j_2+1)\text{k}^2} \\
\sum_{j_{12}j'_{12}ll'J\varepsilon_I}^{*} (2J+1)
|\delta_{vjl,v'j'l'} 
- S^{J\varepsilon_I\varepsilon_P}_{vjl,v'j'l'}(E_c)|^{2}.
\end{multline}
For collisions between para-H$_2$ molecules (nuclear spin $I=0$), $W^+=1, W^-=0$ 
so that only the $\varepsilon_P=+1$ exchange-permutation symmetry is needed 
in the close-coupling calculations, while
for collisions between ortho-H$_2$ molecules (nuclear spin $I=1$), $W^+=2/3, W^-=1/3$ 
and both exchange-permutation symmetries $\varepsilon_P=\pm 1$ are needed.
The total inelastic cross section is the sum of all inelastic
state-to-state cross sections.
The state-to-state rate coefficients are obtained by 
integrating the corresponding cross sections over the Maxwell--Boltzmann distribution of
velocities,
\begin{multline}
k_{v_1j_1v_2j_2 \to v'_1j'_1v'_2j'_2}(T)
= \frac{1}{k_B T} \left( \frac{8}{\pi \mu k_B T} \right)^{1/2}  \\
\int_0^{\infty} \sigma_{v_1j_1v_2j_2 \to v'_1j'_1v'_2j'_2}(E_c) \, e^{-E_c / (k_B T)} \, E_c \, dE_c.
\label{rate-constant}
\end{multline}

\section{Results and discussion}

\subsection{Computational details}

\begin{figure}[b]
\begin{center}
\includegraphics*[width=8cm,keepaspectratio=true]{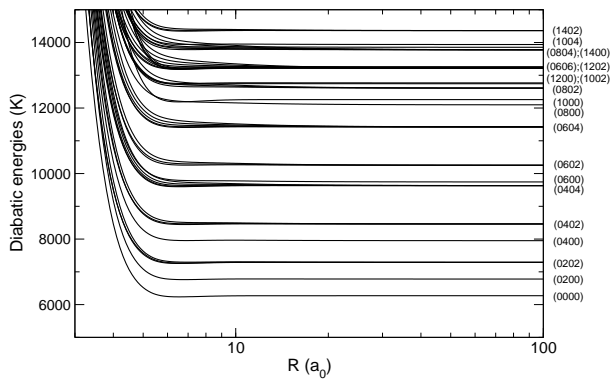}
\caption{
The effective potentials, $V_{\text{eff}}^{J}$,
as functions of $R$ for the H$_4$ system for $J=0, \varepsilon_I=+1, \varepsilon_P=+1$.
\label{SPAG-FIG}}
\end{center}
\end{figure}

In this study  we investigate the scattering of two identical para-H$_2$
molecules 
using the BMKP PES~\cite{Boothroyd02}.
For identical molecule collisions,
the rovibrational quantum numbers obey the ``well ordered states'' 
classification~\cite{Takayanagi65,Green75,Alexander77}.
For convenience we use the term ``combined molecular state" (CMS),
for a combination of 
two rovibrational states of two H$_2$ molecules. In this notation, 
H$_2(v_1,j_1)$ + H$_2(v_2,j_2)$ collisions will
be denoted as $(v_1 j_1 v_2 j_2)$ and its  energy is given by
$\varepsilon_{vj} = \varepsilon_{v_1j_1} + \varepsilon_{v_2j_2}$.
The CMS represents 
an unique quantum state of the diatom - diatom system
before or after the collision. 
To gain physical insights into the scattering
process and also to roughly estimate the number of CMSs to be included in the 
scattering calculations, we plot in Fig.~\ref{SPAG-FIG}
the effective potentials which are composed of the
diabatic energies and the centrifugal terms,
\begin{eqnarray}
V_{\text{eff}}^{J}(R)=
\varepsilon_{vj} 
+ {\cal U}^{JM\varepsilon_I \varepsilon_P}_{vjl,vjl}(R)
+ \frac{\hbar^2 l(l+1)}{2 \mu R^2} ,
\label{DIABNRJ}
\end{eqnarray}
as functions of the radial coordinate, $R$,
evaluated using the BMKP PES.
At large separations, the energies of the
different potential curves converge to that of the corresponding CMSs as indicated by
the numbers $(v_1 j_1 v_2 j_2)$ in Fig.~\ref{SPAG-FIG}. The energy of the CMS (0000) 
is 6270.73~K and that of (0200) is 6780.35~K. For the vibrationally excited case,
(1000), the energy is 12257.57~K. The energy is relative to 
the minimum of the H$_2$ potential.
Since the interaction potential between two H$_2$ molecules 
is relatively weak 
the density of CMSs is rather sparse.
The number of diabatic channels included in the calculation 
depends on the energy. For collision energies $E_c < 100 $~K,
a cut-off energy is used to restrict the number of channels. This
allows one to include all rotational and vibrational levels below the cut-off energy in 
the scattering calculations. 
We used cut-off energies of
20142.76~K (14000~cm$^{-1}$)
for $J=0-3$, and 15826.45~K (11000~cm$^{-1}$)
for $J=4-10$ for ultralow energy calculations. 
However, this approach may lead to a proliferation in the
number of channels at higher energies as the cut-off energy needs to be higher.
Thus only a restricted set of rovibrational levels are included in the calculations.
In the present work we retained all
CMSs
with quantum numbers
$v=0,j=0-8$,
and $v=1,j=0,2$ for calculations at higher energies.
The close-coupling equations~\eqref{CC2} were propagated using the  
log-derivative matrix propagation method of Johnson~\cite{Johnson73}
and Manolopoulos~\cite{Manolopoulos86} to an
asymptotic matching distance of $R=53 \ a_0$.
For all collision energies, the inelastic cross section
is converged at this value of the matching distance.
However, to get converged elastic cross section
for ultralow energies $E_c < 1$~K,
we propagate the elastic channel from $R=53 \ a_0$
to a sufficiently large distance, at which
the interaction potential $U$ in Eq.~\eqref{Potential} is less than $10^{-5} \times E_c$.

\subsection{Vibrational relaxation in H$_2(v=1,j=0)$ + H$_2(v=0,j=0)$ collisions}

\begin{figure}[t]
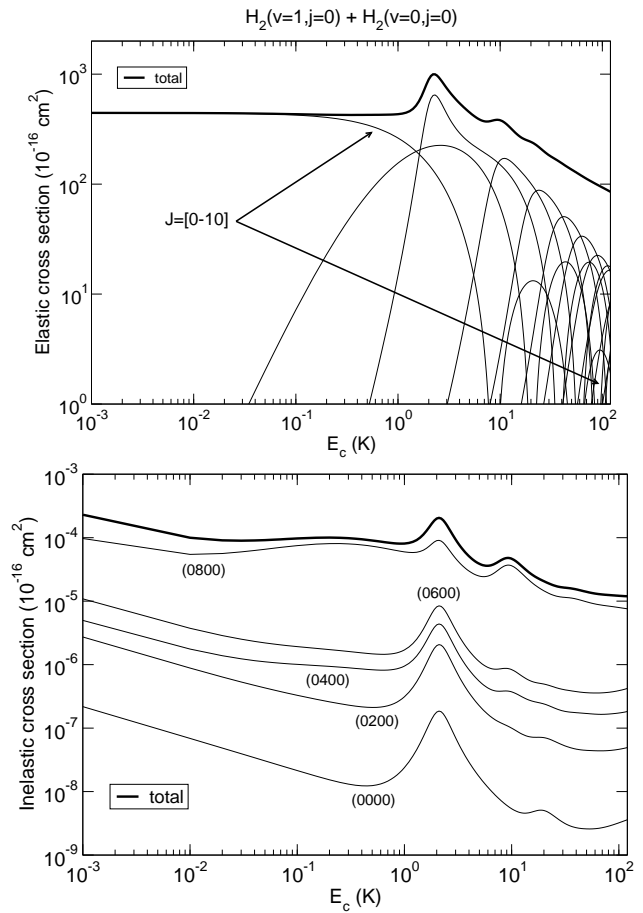

\begin{center}
\includegraphics*[height=6cm,keepaspectratio=true]{XSEL_H2-H2-allJ-vj1000-1nKto120K-BW.eps} \\
\includegraphics*[height=6cm,keepaspectratio=true]{XSIN_H2-H2-allJ-vj1000-1nKto120K-rotdist.eps} 
\caption{
Upper panel: Partial-wave resolved and total elastic cross sections
for H$_2(v=1,j=0)$ + H$_2(v=0,j=0)$ collisions
as a function of the collision energy.
Lower panel: Inelastic state-to-state cross sections
for the H$_2(v=1,j=0)$ + H$_2(v=0,j=0)$ system
as a function of the collision energy.
For clarity, only the final CMSs
$(0j00)$ have been shown.
\label{XSEL-ULE-1000-FIG}}
\end{center}
\end{figure}

In the upper panel of Fig.~\ref{XSEL-ULE-1000-FIG} we show the $J$-resolved partial cross sections
and their sum for the elastic channel for collision energies ranging from
$10^{-3}-120$~K. Though only the $s$-wave contributes at
energies below $10^{-2}$ K, the partial wave sum requires up to $J=10$ at $E_c=100$~K
to yield converged cross sections.
The state-to-state  inelastic cross sections and their sum are shown 
in the lower panel of Fig.~\ref{XSEL-ULE-1000-FIG} 
for H$_2(v=1,j=0)$ + H$_2(v=0,j=0)$ collisions 
for $E_c=10^{-3}-120$~K.
The cross sections of the inelastic processes are much smaller than that
of elastic scattering. Due to the small anisotropy of 
the interaction potential with respect to the stretching of the molecule, the interaction
potential matrix elements responsible for  vibrational
relaxation channels are quite small compared to the isotropic part responsible
for elastic scattering. Thus, the couplings between different CMSs
are rather weak and state-to-state transitions are less probable than  elastic scattering.
The diabatic potential curves in Fig.~\ref{SPAG-FIG} shows that
the density of states is quite small and the energy levels are
sparsely spaced leading to small inelastic couplings.
This is in contrast with the heavier alkali-metal systems with very high density of states,
for which elastic scattering is less important than
inelastic collisions at ultralow collision energies~\cite{Quemener07}. 
In Fig.~\ref{XSEL-ULE-1000-FIG},
the CMS (0800) is the most probable final state. This corresponds to a case
where the vibrationally excited molecule H$_2(v=1,j=0)$ relaxes
to the rotationally excited H$_2(v=0,j=8)$ state upon collisions with
the ground state molecule H$_2(v=0,j=0)$. The final state distribution is
determined by a compromise between conservation of the total internal energy and the rotational
angular momenta of the colliding molecules.

\begin{figure}[t]
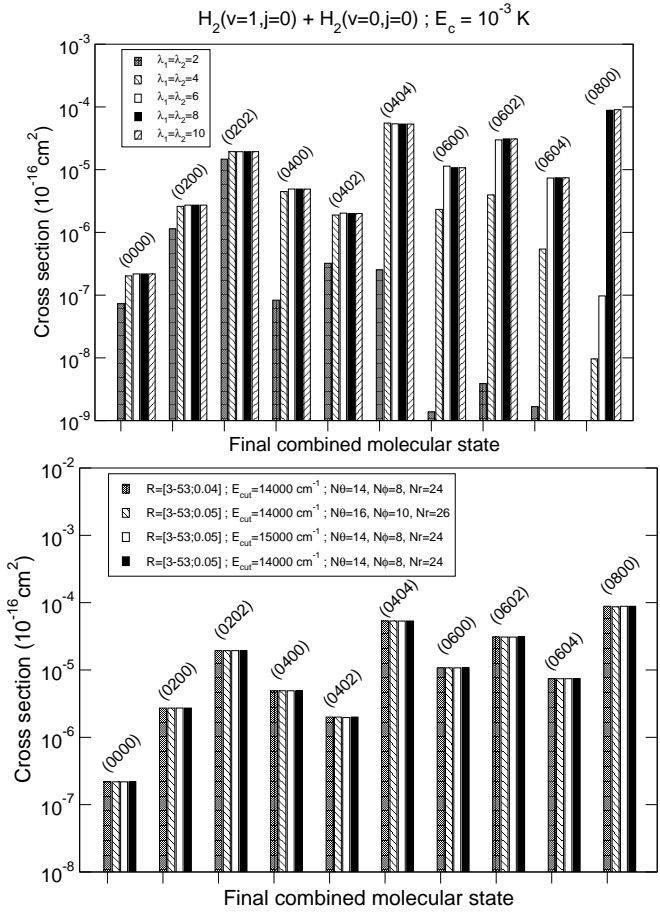

\begin{center}
\includegraphics*[height=6cm,keepaspectratio=true]{XS-ST2ST_H2-H2-vj1000-CONVERGENCE-fig1-1mK.eps} \\
\includegraphics*[height=6cm,keepaspectratio=true]{XS-ST2ST_H2-H2-vj1000-CONVERGENCE-fig2-1mK.eps}
\caption{
State-to-state cross sections for H$_2(v=1,j=0)$ + H$_2(v=0,j=0)$
collisions at an energy of 10$^{-3}$~K.
The parameters corresponding to the black distributions are adopted in the final production
calculations.
Upper panel: convergence with respect to the angular anisotropic terms of the PES.
Lower panel: convergence with respect to the number of points for the radial
and angular integrations and the  cut-off energy for the basis set.
\label{XS-STTOST-FIG}}
\end{center}
\end{figure}

In Fig.~\ref{XS-STTOST-FIG} we present the state-to-state rovibrational  populations of
the two molecules in  H$_2(v=1,j=0)$ + H$_2(v=0,j=0)$  collisions
at an energy of $E_c=10^{-3}$~K. Since rotational level resolved 
state-to-state cross sections are more
difficult to converge compared to initial state-selected cross sections, we 
also include results from our convergence studies. Figure~\ref{XS-STTOST-FIG} shows that the
results are robust and converged with respect to all numerical parameters
employed in the calculations.
The upper panel shows cross sections for different 
values of $\lambda_1=\lambda_2$ in Eq.~\eqref{POTLAMBDA}.
It is seen that $\lambda_1=\lambda_2=8$ or higher is needed to accurately
calculate cross sections for the dominant final CMS, (0800).
The convergence of the results with respect to the cut-off energy,
the size $\Delta R$ of the radial interval,
the number of DVR points 
$N_{r_1}=N_{r_2} \equiv N_{r},N_{\theta_1}=N_{\theta_2} \equiv N_{\theta}$, and $N_{\phi}$
used for the vibrational and rotational wave functions
are presented in the lower panel. These results are obtained with 
$\lambda_1=\lambda_2=8$ 
for the angular expansion of the interaction potential. 
For the results reported in the final production
calculations
for the BMKP PES
we used
$\lambda_1=\lambda_2=8$, 
$\Delta R = 0.05 \ a_0$, 
$N_{\theta}=14, ~ N_{\phi}=8$, and $N_{r}=24$.

\begin{figure}[t]
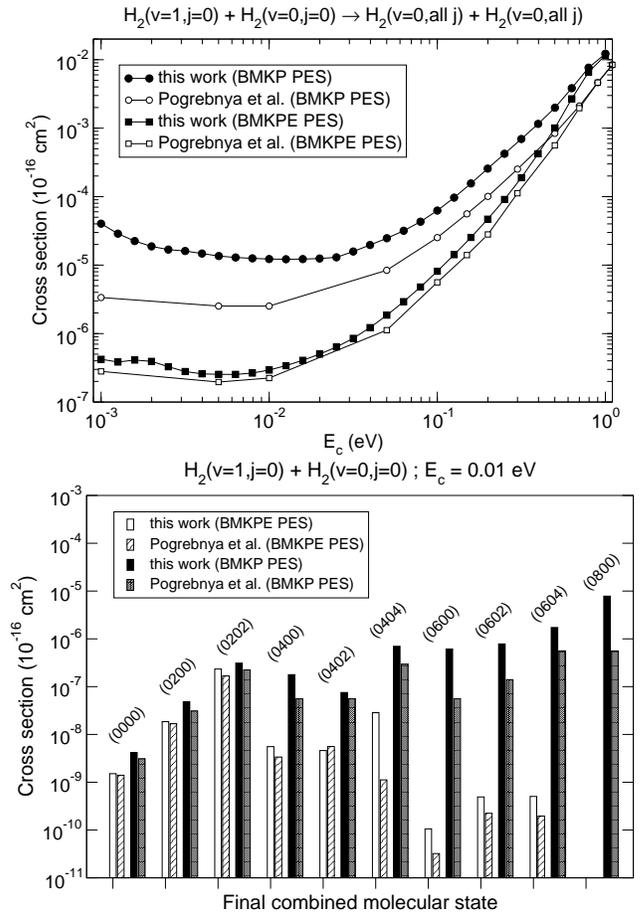

\begin{center}
\includegraphics*[height=6cm,keepaspectratio=true]{XSELIN_H2-H2-v1j8-allJ-ecut10000-vj1000-tov0-0eVto1eV-ANG-BW.eps} \\
\includegraphics*[height=6cm,keepaspectratio=true]{XS-ST2ST_H2-H2-vj1000-ANG-0.01eV.eps}
\caption{
Upper panel: Vibrational relaxation cross sections as functions of the collision energy
for the H$_2(v=1,j=0)$ + H$_2(v=0,j=0)$ system.
Lower panel: State-to-state cross sections at a collision energy of $E_c=0.01$~eV.
\label{XSIN-THERMAL-1000-FIG}}
\end{center}
\end{figure}

Figure~\ref{XS-STTOST-FIG} shows that vibrational relaxation of
H$_2(v=1,j=0)$ by collisions with H$_2(v=0,j=0)$ is driven by
 high order anisotropic terms of the BMKP PES.
Pogrebnya et al.~\cite{Pogrebnya02} had previously
investigated this issue and they had found
that the high anisotropic terms of the 
BMKP PES 
lead to large values of the vibrational relaxation rate coefficients
compared to experimental results.
They found that a restricted version of this PES, referred to as the BMKPE PES
which includes only $\lambda_1=\lambda_2=2$ components in  Eq.~\eqref{POTLAMBDA}
for the angular expansion of the interaction potential, yields better results
in comparison with experimental data.
In the upper panel of Fig.~\ref{XSIN-THERMAL-1000-FIG} we show
the cross section 
for  
H$_2(v=1,j=0)$ + H$_2(v=0,j=0)$ $\to$ H$_2(v=0)$ + H$_2(v=0)$ collisions
for $E_c = 0.001 - 1$~eV evaluated using 
the BMKPE (filled squares) and BMKP (filled circles) PESs.
The corresponding results of Pogrebnya et al.~\cite{Pogrebnya02}
are also shown (open squares and circles) for comparison.
While the results on the BMKPE PES
are in overall good agreement, the results on the BMKP PES differ
significantly at low collision energies (0.001~eV).
The agreement improves with increase in collision energy and both cross sections show 
similar energy dependence for $E_c=0.1-1$~eV. 
In the lower panel of Fig.~\ref{XSIN-THERMAL-1000-FIG}, 
we compare state-to-state cross sections 
at $E_c=0.01$~eV for the BMKPE (white) and BMKP (black) PESs 
with the corresponding results of Pogrebnya et al. (dashed or gray).
Our results
on both PESs 
are in good agreement
with those of Pogrebnya et al.
for the CMSs (0000), (0200), (0202), (0400), (0402)
but
differ 
for the CMSs (0404), (0600), (0602), (0604), (0800).
We believe that the  differences are due to the CSA method
employed in the work of Pogrebnya et al. This applies to the 
results in both panels of Fig.~\ref{XSIN-THERMAL-1000-FIG}. 
The maximum value of the total angular momentum projection quantum number, $\Omega$,
on the body-fixed axis $\vec{R}$
used in the calculations of Pogrebnya et al. is $|\Omega|_{\text{max}}=4$.
In a body-fixed frame,
the orbital angular momentum $\vec{l}$
is always perpendicular to the body-fixed axis $\vec{R}$ and the value $m_l$ of 
its projection on $\vec{R}$
is zero. 
This implies that the projection of $\vec{J} = \vec{j}_{12} + \vec{l}$
on the body-fixed axis is $\Omega = m_{j_{12}}$.
If $|\Omega|_{\text{max}}$ is restricted to 4,
so does $|m_{j_{12}}|$. This is equivalent to exclude all projection quantum numbers
$|m_{j_{12}}| > 4$. Since no such restriction is
imposed in our calculations, the differences
between the exact and CSA methods
for CMSs with quantum numbers $j_{12} > 4$ and $|m_{j_{12}}| > 4$, 
such as (0404), (0600), (0602), (0604), (0800),
are attributed to the restriction on $\Omega$ imposed in the CSA calculations.
The good agreement between the full quantum and CSA  methods
for the BMKPE PES is attributed to the negligible contributions of
the final states (0404), (0600), (0602), (0604) and (0800). Since the restriction on $\Omega$
primarily impacts these final states, it does not significantly affect the
cross sections computed using the BMKPE PES.

\begin{figure}[h]
\begin{center}
\includegraphics*[height=6cm,keepaspectratio=true]{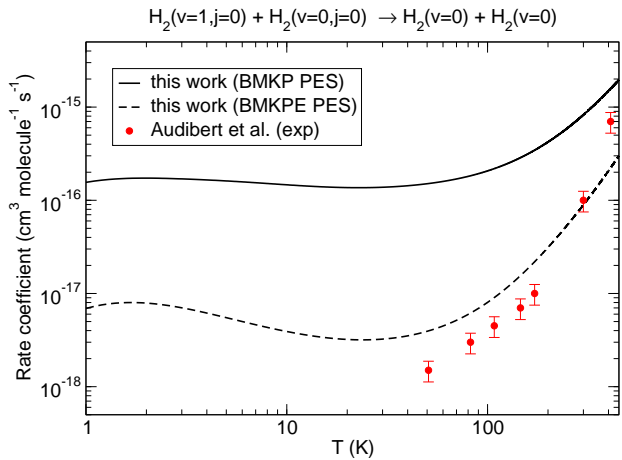}
\caption{
(Color online) Rate coefficient for the vibrational relaxation of H$_2(v=1,j=0)$
in collisions with H$_2(v=0,j=0)$   as a function of the temperature.
Results obtained using the BMKP and BMKPE PESs are compared with the experimental results of
Audibert et al.~\cite{Audibert75}.
\label{RATE-1000-FIG}}
\end{center}
\end{figure}

The rate coefficients for the vibrational relaxation of
H$_2(v=1,j=0)$ in collision with 
H$_2(v=0,j=0)$
as  functions of the temperature
are reported in Fig.~\ref{RATE-1000-FIG}
for the BMKP PES (solid curve) and the BMKPE PES (dashed curve).
The experimental results of Audibert et al.~\cite{Audibert75}
are also shown for comparison.
For the reasons discussed above, the BMKP PES predicts
rate coefficients that are 
about two orders of magnitude larger
than the experimental results for $T<200$~K.
The less anisotropic BMKPE PES yields results in better agreement with the
experimental results. The agreement is very good at $T=300$~K. At lower temperatures,
the calculations predict slightly larger values than the 
experimental results.
For 
$T > 300$~K, the experimental
result is higher than the theoretical results on the BMKPE PES.

\subsection{Rotational excitation and de-excitation in H$_2(v=0,j=0,2)$ + H$_2(v=0,j=0,2)$ collisions}

\begin{figure}[t]
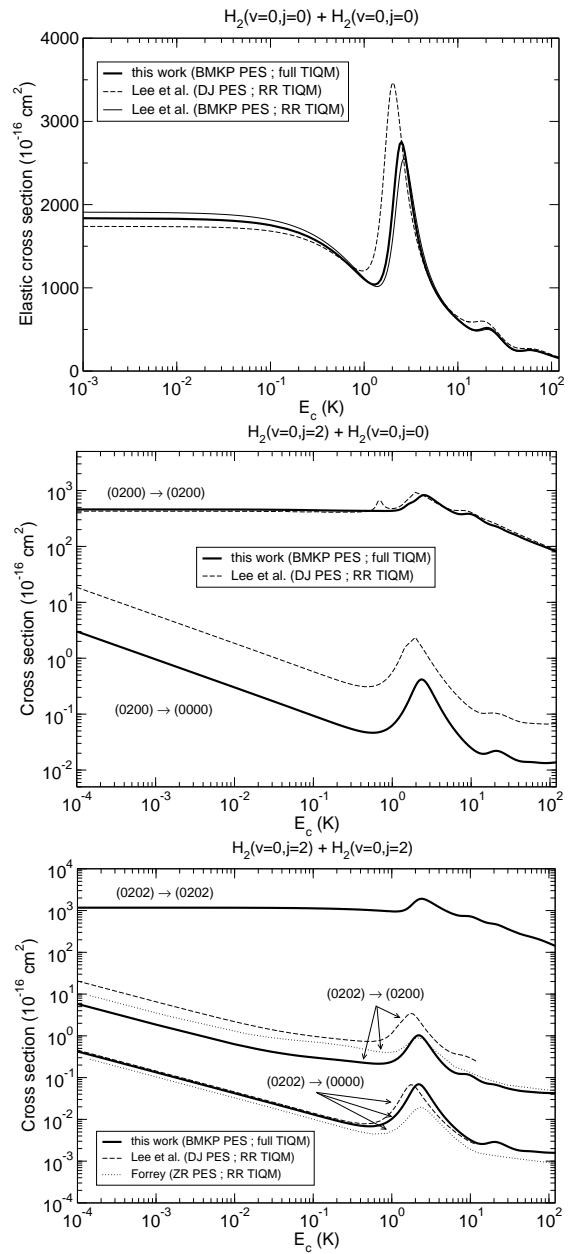

\begin{center}
\includegraphics*[height=5.5cm,keepaspectratio=true]{XSEL_H2-H2-allJ-vj0000-1nKto120K.eps}
\includegraphics*[height=5.5cm,keepaspectratio=true]{XSELIN_H2-H2-allJ-vj0002-1nKto120K.eps}
\includegraphics*[height=5.5cm,keepaspectratio=true]{XSELIN_H2-H2-allJ-vj0202-1nKto120K.eps}
\caption{
Elastic and inelastic cross sections as functions of the collision energy.
Upper panel: Elastic cross sections for  H$_2(v=0,j=0)$ + H$_2(v=0,j=0)$ collisions.
Middle panel: Elastic and inelastic cross sections
for H$_2(v=0,j=2)$ + H$_2(v=0,j=0)$  collisions.
Lower panel: Elastic and inelastic cross sections
for  H$_2(v=0,j=2)$ + H$_2(v=0,j=2)$ collisions.
\label{XSEL-ULE-0000-0200-0202-FIG}}
\end{center}
\end{figure}

In Fig.~\ref{XSEL-ULE-0000-0200-0202-FIG} we show 
cross sections for 
H$_2(v=0,j=0)$ + H$_2(v=0,j=0)$ (upper panel),
H$_2(v=0,j=2)$ + H$_2(v=0,j=0)$ (middle panel)
and H$_2(v=0,j=2)$ + H$_2(v=0,j=2)$ (lower panel) collisions on the BMKP PES. 
Previous results of Lee et al.~\cite{Lee06} on the BMKP and DJ PESs and Forrey~\cite{Forrey01}
on the ZR PES based on the rigid rotor 
model are also shown for comparison.
The results on the BMKP PES in the upper panel
show that the rigid rotor approximation
is reliable in predicting elastic cross sections for collisions between 
two ground  state  H$_2$
molecules. The DJ potential predicts similar results as the BMKP potential 
except for the slight shift in the location of the resonance at about $E_c=2$~K. The 
agreement between the DJ and BMKP results is excellent for collision energies above 4~K. 
Overall, the elastic cross section appears to be less sensitive to the choice of 
the PES and the dynamics approximation. 
In the middle panel of Fig.~\ref{XSEL-ULE-0000-0200-0202-FIG} we show 
the elastic and inelastic rotational de-excitation cross sections for
H$_2(v=0,j=2)$ + H$_2(v=0,j=0)$ collisions. As in the case of collisions 
between ground state molecules, the elastic cross sections from rigid rotor calculations
of Lee et al.~\cite{Lee06} using the DJ PES and the present full-dimensional
results on the
BMKP surface are in good agreement with each other. 
However, the inelastic cross sections show
significant differences.
The inelastic rotational de-excitation cross section on the DJ PES
is about an order of magnitude
larger than that derived from the BMKP PES. This confirms conclusions derived
from previous calculations~\cite{Lee06,Otto08} that the  BMKP PES is not reliable in predicting accurate
values of rotational excitation cross sections, especially for the $\Delta j=2$ rotational
transitions.
In the lower panel we show elastic and inelastic rotational de-excitation cross sections for
H$_2(v=0,j=2)$ + H$_2(v=0,j=2)$ collisions. 
Interestingly, (0202) $\to$ (0000) inelastic cross sections are comparable
in magnitude for the BMKP PES, the DJ PES, and also for the ZR PES. 
However, for the (0202) $\to$ (0200) transition, the BMKP potential
yields smaller cross sections than DJ and ZR potentials for collision energies below 2~K. 
The agreement improves for higher collision energies, especially with that from the 
ZR potential.

\begin{figure}[h]
\begin{center}
\includegraphics*[height=12cm,width=8cm,keepaspectratio=true]{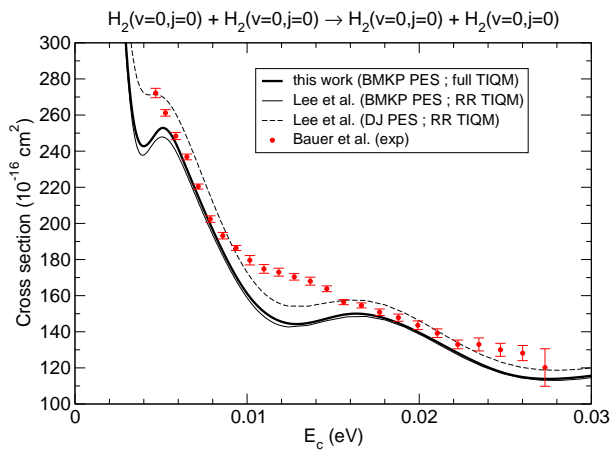}
\caption{
(Color online) Elastic cross section
for  H$_2(v=0,j=0)$ + H$_2(v=0,j=0)$
collisions as a function of the incident collision energy. The experimental results
of Bauer et al.~\cite{Bauer76} and theoretical results of Lee et al.~\cite{Lee06}
are also shown. 
\label{XS-EL-FIG}}
\end{center}
\end{figure}

The elastic cross section of the H$_2(v=0,j=0)$ + H$_2(v=0,j=0)$
collision is shown in Fig.~\ref{XS-EL-FIG}
for collision energies up to 0.03~eV. The theoretical results of Lee et al.~\cite{Lee06} and
 experimental results of Bauer et al.~\cite{Bauer76}
are also included for comparison.
The present results 
do not differ significantly from the results
of Lee et al.
The full calculation predicts slightly larger values for the 
elastic cross section compared to the rigid rotor results.
Overall, the DJ PES predicts results in better agreement
with the experimental results.

\begin{figure}[h]
\begin{center}
\includegraphics*[height=12cm,width=8cm,keepaspectratio=false]{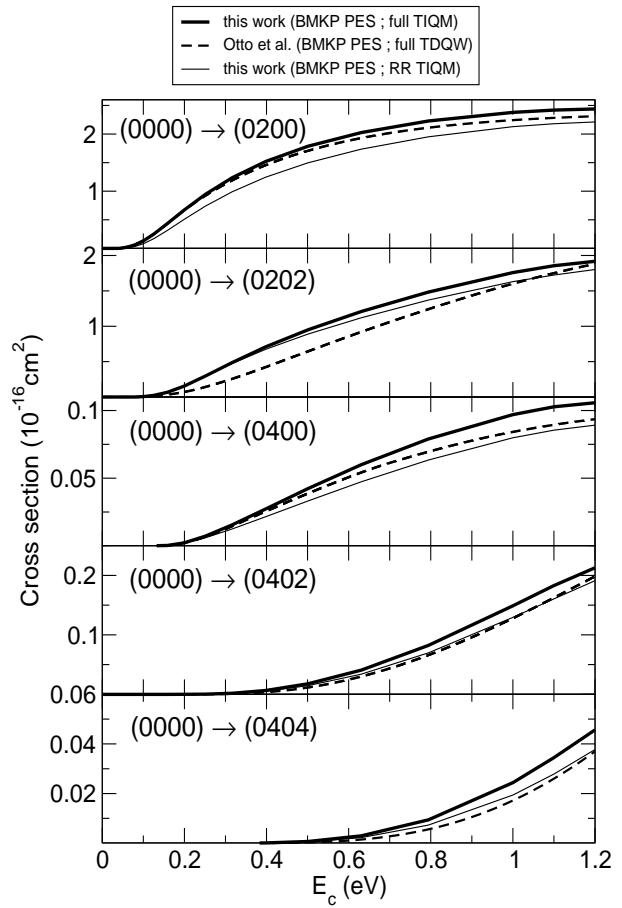}
\caption{
State-to-state cross sections  for rotational excitation
in H$_2(v=0,j=0)$ + H$_2(v=0,j=0)$ collisions as functions of the collision energy. The
results of Otto et al.~\cite{Otto08} are included for comparison.
\label{ST2STXS-THERMAL-0000-FIG}}
\end{center}
\end{figure}

A number of recent studies have reported rotational
excitation cross sections in collisions between two rovibrationally ground state
H$_2$ molecules~\cite{Gatti05,Lee06,Otto08}. The studies of Gatti et al.~\cite{Gatti05}
and Otto et al.~\cite{Otto08} employed the full-dimensional multiconfiguration time-dependent Hartree (MCTDH)
approach while that of Lee et al.~\cite{Lee06} employed the TIQM approach
within the rigid rotor approximation. Though the agreement between these calculations
were generally good, important differences were found for some transitions, especially
for (0000) $\to$ (0202) and (0000) $\to$ (0404) transitions. In 
Fig.~\ref{ST2STXS-THERMAL-0000-FIG} we show
rotational excitation cross sections in
H$_2(v=0,j=0)$ + H$_2(v=0,j=0)$ collisions leading to
final CMSs (0200), (0202), (0400), (0402), and (0404). To obtain converged
cross sections, these calculations
include contributions from $J=0-100$. Also
included are TIQM rigid rotor results
from the present work, and full-dimensional
time-dependent calculations of Otto et al.~\cite{Otto08}. Figure~4 of Ref.~\cite{Otto08}
presents similar comparison with the results of Lee et al.~\cite{Lee06}, Lin et al.~\cite{Lin02}, 
and Sultanov et al.~\cite{Sultanov06a}.
Figure~\ref{ST2STXS-THERMAL-0000-FIG} shows that our results are generally in good 
agreement with those of Otto et al. for the (0000) $\to$ (0200) and 
(0000) $\to$ (0400) transitions. The agreement is better for collision energies
lower than 0.5 eV. The rigid rotor
calculations from the present study
underestimate  the cross sections.
The largest discrepancy between the present work and the time-dependent
calculations occur for the (0000) $\to$ (0202) transition. 
Overall, for all of the transitions shown in Fig.~\ref{ST2STXS-THERMAL-0000-FIG}
the agreement with the time-dependent results is better at low collision energies and
it deteriorates with increasing collision energy. 
This also applies to the rigid rotor results compared to the full-dimensional
calculations from the present study. 
Our full-dimensional
calculations include $v=0, j=0-8$ and $v=1,j=0-2$ levels in the basis set. It is possible that
a larger basis set with additional rotational levels would yield results in
better agreement with the time-dependent results at higher energies. 
However, this
does not explain the discrepancy between the time-dependent and time-independent
calculations for the (0000) $\to$ (0202) transition.
Calculations using significantly larger basis sets 
are computationally demanding and beyond the scope of this paper.

\begin{figure}[h]
\begin{center}
\includegraphics*[height=12cm,width=8cm,keepaspectratio=true]{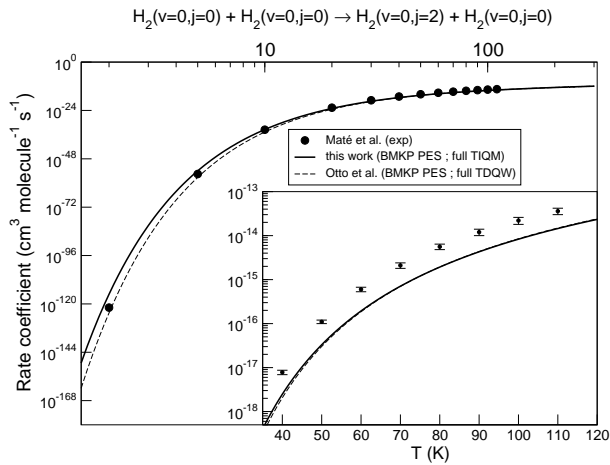}
\caption{
Rate coefficient as a function of the temperature
for  H$_2(v=0,j=0)$ + H$_2(v=0,j=0)$ $\to$ H$_2(v=0,j=2)$ + H$_2(v=0,j=0)$ collisions.
The TDWP results of Otto et al.~\cite{Otto08} and experimental 
results of Mat{\'e} et al.~\cite{Mate05} are also included for comparison.
The TDWP results are almost identical
to the present results
for $T > 10$~K.
\label{RATE-THERMAL-FIG}}
\end{center}
\end{figure}

In Fig.~\ref{RATE-THERMAL-FIG}
we show the temperature dependence of the rate coefficient
for 
H$_2(v=0,j=0)$ + H$_2(v=0,j=0)$ $\to$ H$_2(v=0,j=2)$ + H$_2(v=0,j=0)$
collisions along with the corresponding
experimental results of Mat{\'e} et al.~\cite{Mate05}
and the TDQW results of Otto et al.~\cite{Otto08}.
It has previously been shown that for this transition
the BMKP PES underestimates the rate 
coefficient by about an order of magnitude \cite{Lee06,Otto08}.
Our results confirm this finding. As Fig.~\ref{RATE-THERMAL-FIG}
illustrates, our results are in very good agreement with those of 
Otto et al.~\cite{Otto08} although the corresponding
cross sections given in Fig.~\ref{ST2STXS-THERMAL-0000-FIG} show some differences
at collision
energies above 0.5~eV. This is because in the temperature range of $10-300$~K,
the main contribution to the rate coefficients occurs from
cross sections at energies below 0.5~eV.

\section{Conclusion}

In this paper we present explicit quantum scattering
calculations of 
collisions between two para-H$_2$ molecules
in $v=0$ and $v=1$ vibrational levels
from ultralow to thermal energies. A 
new code~\cite{Roman06} based on a full-dimensional time-independent quantum
scattering method has been employed for the calculations. 
Using this code we have been able to test the reliability of the
coupled-states approximation for accurately predicting vibrational relaxation 
rate coefficients for H$_2(v=1,j=0)$ + H$_2(v=0,j=0)$ collisions. Comparison
of our results with previous coupled-states 
calculations of Pogrebnya and Clary shows that the coupled-states
approximation is not very reliable for this system in predicting
vibrational relaxation cross sections at low energies. This is in part due to
the high order angular anisotropic terms of the H$_4$ BMKP PES.
We obtained
good agreement with the results of Pogrebnya and Clary for the BMKPE potential
which excludes the high order anisotropic terms of the interaction potential.
The relaxation rate coefficient is found to be generally in good agreement
with experimental results when the BMKPE PES is used.

We explored pure rotational transitions in
H$_2(v=0,j=0,2)$ + H$_2(v=0,j=0,2)$ collisions at ultralow energies
and
compared our results
with 
those of Forrey and Lee et al., using different PESs.
While the inelastic processes are sensitive to the PES employed,
the elastic processes do not show significant differences. 
We also presented pure rotational excitation
of H$_2(v=0,j=0,2)$ + H$_2(v=0,j=0,2)$
and compared our results with those of Otto et al.
The TIQM
and TDWP methods give comparable results for collision energies up to 0.5~eV.
But some discrepancies exist for (0000) $\to$ (0202)
and (0000) $\to$ (0404) rotational transitions, presumably due to
the contributions from the $v=1$ level.

Future works will involve ortho-para and ortho-ortho collisions
of hydrogen molecules as well as H$_2-$HD, D$_2-$D$_2$ and HD$-$HD collisions.

\section{Acknowledgments}

This work was supported  by NSF grants \# PHY-0555565, ATM-0635715, 
AST-0607524, and NASA grant NNG06GC94G.
We thank R. V. Krems for initial collaboration on this project and for many stimulating discussions and
consultations. We thank T.-G. Lee and F. Otto for providing us their theoretical results.


\begin{thebibliography}{53}

\bibitem{Bahns00}
J. T. Bahns, W. Stwalley, and P. L. Gould,
Adv. At. Mol. Opt. Phys. {\bf 42}, 171 (2000).

\bibitem{Masnou01}
F. Masnou--Seeuws and P. Pillet,
Adv. At. Mol. Opt. Phys. {\bf 47}, 53 (2001).

\bibitem{Bethlem03}
H. L. Bethlem and G. Meijer,
Int. Rev. Phys. Chem. {\bf 22}, 73 (2003).

\bibitem{Doyle04}
J. Doyle, B. Friedrich, R. V. Krems, and F. Masnou--Seeuws,
Eur. Phys. J. D {\bf 31}, 149 (2004).

\bibitem{Hutson06}
J. M. Hutson and P. Sold\'{a}n,
Int. Rev. Phys. Chem. {\bf 25}, 497 (2006).

\bibitem{Hutson07a}
J. M. Hutson and P. Sold\'{a}n,
Int. Rev. Phys. Chem. {\bf 26}, 1 (2007).

\bibitem{roman-review}
R. V. Krems,
Int. Rev. Phys. Chem. {\bf 24}, 99 (2005).

\bibitem{Roman-pccp}
R. V. Krems,
Phys. Chem. Chem. Phys. {\bf 10}, 4079 (2008).

\bibitem{bala-cpl-2001}
N. Balakrishnan and A. Dalgarno,
Chem. Phys. Lett. {\bf 341}, 652 (2001).

\bibitem{weck06} 
P. F. Weck and N. Balakrishnan,
Int. Rev. Phys. Chem. {\bf 25}, 283 (2006).

\bibitem{Quemener08b}
G. Qu{\'e}m{\'e}ner and N. Balakrishnan,
J. Chem. Phys. {\bf 128}, 224304 (2008).

\bibitem{bodo-review}
E. Bodo and F. A. Gianturco,
Int. Rev. Phys. Chem. {\bf 25}, 313 (2006).

\bibitem{Soldan02}
P. Sold\'{a}n, M. T. Cvita\v{s}, J. M. Hutson, P. Honvault, and J.-M. Launay,
Phys. Rev. Lett. {\bf 89}, 153201 (2002).

\bibitem{Quemener04}
G. Qu\'{e}m\'{e}ner, P. Honvault, and J.-M. Launay,
Eur. Phys. J. D {\bf 30},
201 (2004).

\bibitem{Quemener05}
G. Qu\'{e}m\'{e}ner, P. Honvault, J.-M. Launay, P. Sold\'{a}n, D. E. Potter,
and J. M. Hutson,
Phys. Rev. A {\bf 71}, 032722 (2005).

\bibitem{Cvitas05a}
M. T. Cvita\v{s}, P. Sold\'{a}n, J. M. Hutson, P. Honvault, and J.-M. Launay,
Phys. Rev. Lett. {\bf 94}, 033201 (2005).

\bibitem{Cvitas05b}
M. T. Cvita\v{s}, P. Sold\'{a}n, J. M. Hutson, P. Honvault, and J.-M. Launay,
Phys. Rev. Lett. {\bf 94}, 200402 (2005).

\bibitem{Quemener07}
G. Qu\'{e}m\'{e}ner, J.-M. Launay, and P. Honvault,
Phys. Rev. A {\bf 75}, 050701(R) (2007).

\bibitem{Quemener09}
G. Qu\'{e}m\'{e}ner, N. Balakrishnan, and B. K. Kendrick,
arXiv:0811.4377v1 [physics.chem-ph].

\bibitem{Zarur74}
G. Zarur and H. Rabitz,
J. Chem. Phys. {\bf 60}, 2057 (1974).

\bibitem{Schaefer89}
J. Schaefer and W. E. K\"ohler,
Z. Phys. D: At., Mol. Clusters {\bf 13}, 217 (1989).

\bibitem{Diep00}
P. Diep and J. K. Johnson, J. Chem. Phys. {\bf 112}, 4465 (2000).

\bibitem{Patkowski08}
K. Patkowski, W. Cencek, P. Jankowski, K. Szalewicz,
J. B. Mehl, G. Garberoglio, and A. H. Harvey,
J. Chem. Phys. {\bf 129}, 094304 (2008).

\bibitem{Forrey01}
R. C. Forrey, 
Phys. Rev. A {\bf 63}, 051403(R) (2001).

\bibitem{Forrey02}
R. C. Forrey, 
Phys. Rev. A {\bf 66}, 023411 (2002).

\bibitem{Mate05}
B. Mat\'{e}, F. Thibault, G. Tejeda, J. M. Fern\'{a}ndez, and S.
Montero, J. Chem. Phys. {\bf 122}, 064313 (2005).

\bibitem{Lee06}
T.-G. Lee, N. Balakrishnan, R. C. Forrey, P.C. Stancil, D.R. Schultz,
and G. J. Ferland,
J. Chem. Phys. {\bf 125}, 114302 (2006).

\bibitem{Sultanov06a}
R. A. Sultanov and D. Guster, Chem. Phys. {\bf 326}, 641 (2006).

\bibitem{Sultanov06b}
R. A. Sultanov and D. Guster, Chem. Phys. {\bf 428}, 227 (2006).

\bibitem{Schwenke88}
D. W. Schwenke, J. Chem. Phys. {\bf 89}, 2076 (1988).

\bibitem{Aguado94}
A. Aguado, C. Su{\'a}rez, and M. Paniagua,
J. Chem. Phys. {\bf 101}, 4004 (1994).

\bibitem{Boothroyd02}
A. I. Boothroyd, P. G. Martin, W. J. Keogh, and M. J. Peterson,
J. Chem. Phys. {\bf 116}, 666 (2002).

\bibitem{Hinde08}
R. J. Hinde,
J. Chem. Phys. {\bf 128}, 154308 (2008).

\bibitem{Flower98a}
D. R. Flower, 
Mon. Not. R. Astron. Soc. {\bf 297}, 334 (1998).

\bibitem{Flower98b}
D. R. Flower and E. Roueff,
J. Phys. B {\bf 31}, 2935 (1998).

\bibitem{Flower99}
D. R. Flower and E. Roueff,
J. Phys. B {\bf 32}, 3399 (1999).

\bibitem{Flower00a}
D. R. Flower,
J. Phys. B {\bf 33}, L193 (2000).

\bibitem{Flower00b}
D. R. Flower,
J. Phys. B {\bf 33}, 5243 (2000).

\bibitem{Pogrebnya02}
S. K. Pogrebnya and D. C. Clary,
Chem. Phys. Lett. {\bf 363}, 523 (2002).

\bibitem{Pogrebnya03}
S. K. Pogrebnya, M. E. Mandy, and D. C. Clary,
Int. J. Mass. Spectrom. {\bf 223-224}, 335 (2003).

\bibitem{Lin02}
S. Y. Lin and H. Guo, 
J. Chem. Phys. {\bf 117}, 5183 (2002).

\bibitem{Gatti05}
F. Gatti, F. Otto, S. Sukiasyan, and H.-D. Meyer,
J. Chem. Phys. {\bf 123}, 174311 (2005).

\bibitem{Otto08}
F. Otto, F. Gatti, and H.-D. Meyer,
J. Chem. Phys. {\bf 128}, 064305 (2008).

\bibitem{Panda07}
A. N. Panda, F. Otto, F. Gatti, and H.-D. Meyer,
J. Chem. Phys. {\bf 127}, 114310 (2007).

\bibitem{Quemener08a}
G. Qu\'em\'ener, N. Balakrishnan, and R.V. Krems,
Phys. Rev A {\bf 77}, 030704(R) (2008).

\bibitem{Audibert75}
M.-M. Audibert, R. Vilaseca, J. Lukasik, and J. Ducuing,
Chem. Phys. Lett. {\bf 31}, 232 (1975).

\bibitem{Bauer76}
W. Bauer, B. Lantzsch, J. P. Toennies, and K. Walaschewski,
Chem. Phys. {\bf 17}, 19 (1976).

\bibitem{Roman06}
R. V. Krems, TwoBC - quantum scattering program,
University of British Columbia, Vancouver, Canada (2006).

\bibitem{Takayanagi65}
K. Takayanagi, Adv. At. Mol. Phys. {\bf 1}, 149 (1965).

\bibitem{Green75}
S. Green, J. Chem. Phys. {\bf 62}, 2271 (1975).

\bibitem{Alexander77}
M. H. Alexander and A. E. DePristo, 
J. Chem. Phys. {\bf 66}, 2166 (1977).

\bibitem{Arthurs60}
A. M. Arthurs and A. Dalgarno, {Proc.\ R. Soc.} {\bf A 256}, 540 (1960).

\bibitem{Johnson73}
B. R. Johnson, J. Comp. Phys. {\bf 13}, 445 (1973).

\bibitem{Manolopoulos86}
D. E. Manolopoulos, J. Chem. Phys. {\bf 85}, 6425 (1986).

\end{thebibliography}
\end{document}